\documentclass[twocolumn,showpacs,floats,floatfix,superscriptaddress,aps,pra]{revtex4-1}
\usepackage{amsfonts}
\usepackage{amssymb}
\usepackage{amsmath}
\usepackage{calc}
\usepackage{graphicx}
\usepackage{bm}

\usepackage[normalem]{ulem}
\usepackage{amsmath,amssymb}
\usepackage{multirow}
\usepackage{xcolor,soul}
\usepackage{srcltx}
\usepackage{hyperref,graphicx}

\def\be{ \begin{equation}}
\def\ee{ \end{equation}}
\def\bea{ \begin{eqnarray}}
\def\eea{ \end{eqnarray}}
\def\bse{ \begin{subequations}}
\def\ese{ \end{subequations}}
\def\bc{ \begin{center}}
\def\ec{ \end{center}}

\begin{document}

\author{Stefano Longhi} 
\email{stefano.longhi@polimi.it}
\author{Giuseppe Della Valle}
\affiliation{Dipartimento di Fisica, Politecnico di Milano and Istituto di Fotonica e Nanotecnologie del Consiglio Nazionale delle Ricerche, Piazza L. da Vinci 32, I-20133 Milano, Italy}

\title{Optical lattices with exceptional points in the continuum}
  \normalsize


%
\bigskip
\begin{abstract}
\noindent  
The spectral, dynamical and topological properties of physical systems  described by non-Hermitian (including $\mathcal{PT}$-symmetric) Hamiltonians are deeply modified by the appearance of exceptional points and spectral singularities. Here we show that exceptional points in the continuum can arise in 
non-Hermitian (yet admitting and entirely real-valued energy spectrum) optical lattices with engineered defects. At an exceptional point, the lattice sustains a bound state with an energy embedded in the spectrum of scattered states, similar to the von Neumann-Wigner bound states in the continuum of Hermitian lattices. However, the dynamical and scattering properties of the bound state at an exceptional point are deeply different from those of ordinary von Neumann-Wigner bound states in an Hermitian system. In particular, the bound state in the continuum at an exceptional point  is an unstable state that can secularly grow by an infinitesimal perturbation. Such properties are discussed in details for transport of discretized light in a $\mathcal{PT}$-symmetric array of coupled optical waveguides, which could provide an experimentally accessible system to observe exceptional points in the continuum. 
\end{abstract}

\pacs{03.65.-w, 42.82.Et, 42.25.Bs, 11.30.Er, 11.30.Pb}


\maketitle

\section{Introduction}
Non-Hermitian Hamiltonians (NHHs) are widely used to describe open quantum systems in many areas of science \cite{Moiseyev,uffa}.
Interestingly, in certain cases a NHH $\mathcal{H}$ can show an entire real-valued energy spectrum, in spite of non-self-adjointness. Such a remarkable property has been especially investigated for $\mathcal{PT}$ symmetric Hamiltonians \cite{Bender}, although several examples of non-$\mathcal{PT}$ invariant Hamiltonians yet admitting an entire real-valued energy spectrum have been provided. However,  the reality of the spectrum does not correspond to orthogonality of eigenstates and, most importantly, does not ensure
diagonalizability, which may be prevented by the presence
of exceptional points (EPs) in the point spectrum of $\mathcal{H}$ \cite{EP0,EP1,EP2},
or of spectral singularities in the continuous part of the
spectrum \cite{SS}. The physical implications of both EPs and spectral singularities have 
attracted a great attention in recent years and have been investigated in several physical systems \cite{EP1,EP2,caz1,caz2,uff1,uff2,uff3,palma1,palma2,optics1,palma3,SS,SS1}.
Exceptional points correspond to degeneracies
of a NHH where both eigenvalues and eigenvectors of a finite-dimensional Hamiltonian $\mathcal{H}$ coalesce as
a system parameter is varied \cite{EP0,EP1}.  EPs cause ${\mathcal PT}$ symmetry breaking in $\mathcal{PT}$ symmetric systems  of finite dimension. EPs of a NHH exhibit
highly non-trivial characteristics compared with those of
most common Hermitian degeneracies, especially concerning adiabatic features and geometric
phases, which have been demonstrated in a series of experiments using microwave \cite{caz1} and optical \cite{caz2} cavities.  
In the quantum realm, the existence of EPs has
been predicted theoretically in a wide range of systems, such as in atomic or molecular
 spectra \cite{uff1}, in atom waves \cite{uff2}, and in non-Hermitian
Bose-Hubbard models \cite{uff3}. 
Photonic structures in the presence of gain or loss are a
natural arena in which EPs can play a role, since they are
described by a non-Hermitian operator arising
from a complex dielectric function \cite{optics0}.
 Optics has provided in the past few years a formidable testbed where the main features of NHH systems, including those with $\mathcal{PT}$ symmetry, have been experimentally  observed and exploited to mold the flow of light in 
  new ways \cite{optics2}.
Examples of optical NHH admitting EPs include
lasers \cite{palma1}, coupled waveguides \cite{palma2,optics1} and optical resonators \cite{palma3}. \par
Most of previous theoretical and experimental works on EPs have been limited to consider finite dimensional NHH systems, or EPs of resonance states. In infinite-dimensional systems, the appearance of EPs in the continuum of the energy spectrum (not to be confused with spectral singularities \cite{SS,SS1}) has been theoretically studied in few recent works \cite{B1,B2,B3}, mainly on a mathematical perspective. EPs in the continuum are energies $E_0$ embedded in the continuous spectrum of scattered states of $\mathcal{H}$ 
that sustain bound (normalizable) states with a number of associated functions \cite{B2}. Hence at an EP in the continuum the NHH $\mathcal{H}$ supports bound states similar to so-called bound states in the continuum (BIC) of von Neumann-Wigner  type \cite{Wigner} found in the Hermitian case. BIC states have been predicted to occur in a wide range of quantum and classical systems, including atomic and molecular systems  \cite{miscappa1}, semiconductor and mesoscopic structures \cite{miscappa2}, quantum Hall insulators \cite{miscappa3},  and Hubbard models \cite{miscappa4}.  Experimental observations of BIC states in the Hermitian case have been reported in a few recent works using waveguide arrays \cite{palle1,palle1bis} and photonic crystals \cite{palle2}. However, EPs in the continuum are not just BIC states, since they are {\it defective} states \cite{B2}. The main physical implication is that, as opposed to a BIC state in an Hermitian system, the BIC state of an EP in the continuum is an {\it unstable} state, even though the spectrum of $\mathcal{H}$ is entirely real-valued. So far, EPs in the continuum have been predicted to occur for the Schr\"{o}dinger equation with certain specially-tailored complex potentials \cite{B1,B2}, synthesized by application of a double supersymmetric (Darboux) transformation to the free-particle Hamiltonian \cite{susy1}. Such special potentials, besides to show an EP, are also transparent potentials. Unfortunately, they are very difficult to be implemented in any physical system. On the other hand, light transport in discretized optical structures \cite{Longhi} provides a feasible laboratory tool where the physical features of  NHH can be observed \cite{optics1,optics2,optics3}.\par
In this work we introduce EPs in the continuum in discrete (tight-binding) lattices, which can describe light propagation in arrays of evanescently-coupled optical waveguides, and discuss their dynamical and scattering properties. In particular, the different behavior of EPs in the continuum as compared to ordinary BIC modes of Hermitian lattices is highlighted. The  class of discrete optical lattices, with an entirely real-valued energy spectrum and admitting an EP in the continuum, is synthesized by application in a nontrivial way of a double {\it discrete} Darboux transformation \cite{Darb1,Darb2} to a homogenous Hermitian optical lattice.\par The paper is organized as follows. In Sec.II NHH tight-binding lattices with one EP in the continuum are synthesized, using a double Darboux transformation technique. The basic difference between EPs in the continuum and ordinary BIC states of von Neumann-Wigner type is also elucidated. In Sec.III an example of a simple lattice model with modulated hopping rates, which shows an EP at the lattice band center, is presented, and a possible physical realization using an array of evanescently-coupled optical waveguides is suggested. Section IV outlines the main conclusions. Finally,  a few technical issues are discussed in three Appendices.

\section{Tight-binding lattices with exceptional points in the continuum}
A powerful technique to generate  EPs in the continuum for the continuous Schr\"{o}dinger equation is the application of a multiple Darboux (supersymmetric) transformation to the free-particle equation. Multiple supersymmetric transformations have been used  to synthesize either Hermitian \cite{B3,susy1} or non-Hermitian \cite{B1,B2} potentials 
supporting BIC states. In particular, in Ref.\cite{B2} it was shown that the  Schr\"{o}dinger equation $i \partial_t \psi= \mathcal{H} \psi=-\partial^2_x \psi+V(x) \psi$ with the specially-tailored  complex potential $V(x)=16 \alpha^2 [\alpha(x-\lambda) \sin(2 \alpha x)+2 \cos^2(\alpha x)]/[\sin(2 \alpha x)+2\alpha(x-\lambda)]^2$ [with ${\rm Im}(\lambda) \neq0$, $\alpha$ real] sustains the BIC state $\psi_0(x)=\cos(2 \alpha x)/[\sin(2 \alpha x+2 \alpha(x-\lambda)]$ at the energy $E=\alpha^2$, which is an EP in the continuum. Moreover, such a defective potential is invisible, i.e. plane waves with any wave number $k \neq \pm \alpha$ are fully transmitted across the defect with unitary transmittance and no phase delay or advancement, as if the defect were absent. Unfortunately, such specially-tailored complex optical potentials are difficult to be implemented in physical systems like e.g. matter waves or optical systems. On the other hand, discrete optical potentials that describe light transport in  waveguide array structures or optical mesh lattices could provide an experimentally accessible platform to observe such a kind of invisible defects with EPs in the continuum. In this section we aim to synthesize a tight-binding lattice with EPs in the continuum using a discrete analogue of the double Darboux (supersymmetric) transformation. For the sake of clearness, the technique of double Darboux transformation for the discrete Schr\"{o}dinger equation is presented in Appendix A.  

\subsection{Lattice synthesis}
In the nearest-neighbor approximation, a tight binding lattice is described by the Hamiltonian
\begin{equation}
\mathcal{H}= \sum_{n}  \kappa_{n} \left( |n-1\rangle \langle n|+|n
\rangle \langle n-1|\right)  + \sum_n V_n |n \rangle \langle n|
\end{equation}
where $|n\rangle$ is a Wannier state localized at site $n$ of the
lattice, $\kappa_n$ is the hopping rate between sites $|n-1 \rangle$
and $|n \rangle$, and $V_n$ is the energy of Wannier state $|n
\rangle$. The main idea of the the double discrete Darboux technique is to consider an initial homogeneous Hermitian lattice with Hamiltonian $\mathcal{H}=\mathcal{H}_1$, corresponding to $\kappa_n=\kappa$ and $V_n=0$ in Eq.(1), and to synthesize, using the intertwining operator technique and via an intermediate Hamiltonian $\mathcal{H}_2$, a final partner NH lattice Hamiltonian $\mathcal{H}=\mathcal{H}_3$, which is isospectral to $\mathcal{H}_1$ but showing an EP at an energy $E=\mu_1$ embedded in the continuous spectrum.  
The hopping rates and site energies of the intermediate ($\mathcal{H}_2$) and final ($ \mathcal{H}_3$) Hamiltonians are constructed from the eigensolution $\phi^{(1)}_n$ of 
the equation $\mathcal{H}_1 \phi_n^{(1)} =\mu_1 \phi_n^{(1)}$, which is given by
\begin{equation}
\phi_n^{(1)}= \cos(q_0n+ \sigma)
\end{equation}
with $\mu_1=2 \kappa \cos q_0$. In Eq.(2), $q_0$  and $\sigma$ are real-valued parameters, which are chosen such that $\phi^{(1)}_n$ is non-vanishing. Typically, we will assume $q_0$ a rational number, so that there exists a finite number $\epsilon>0$ such that $|\phi^{(1)}_n|> \epsilon$ for any integer $n$. Such a condition ensures that the hopping rates and potential of the intermediate Hamiltonian $\mathcal{H}_2$ are non-singular and bounded. Note that $\mu_1$ belongs to the continuous spectrum of $\mathcal{H}_1$, i.e. it is embedded into the lattice band $ -2 \kappa \leq E \leq 2 \kappa$. For the given choice of $\mathcal{H}_1$, $\mu_1$ and $\phi^{(1)}_n$, a double Darboux transformation can be applied following the procedure outlined in Appendix A. After some lengthy calculations, the following expression for the hopping rates and site potentials of the partner Hamiltonian $\mathcal{H}_3$ can be derived
\begin{equation}
\kappa_n^{(3)}= \kappa \sqrt{\frac{\rho_n \rho_{n-2}}{\rho_{n-1}^2}}
\end{equation}
\begin{eqnarray}
V^{(3)}_n & = & - \kappa \frac{\sin^2 q_0}{\cos(q_0n+ \sigma) \cos[q_0(n-1)+ \sigma]} \nonumber \\
& + & \kappa \frac{\cos[q_0(n-1)+ \sigma]}{\cos(q_0n+ \sigma)} \frac{\rho_{n+1}}{ \rho_{n}}  \\
& -& \kappa \frac{\cos[q_0(n-2)+ \sigma]}{\cos[q_0(n-1)+ \sigma]} \frac{\rho_n}{ \rho_{n-1}} \nonumber
\end{eqnarray}
where we have set
\begin{equation}
\rho_n \equiv \lambda +n + \frac{\sin (q_0 n) \cos [q_0(n-1)+2 \sigma] }{\sin q_0}
\end{equation}
and where $\lambda$ is an arbitrary complex parameter. At the energy $E=\mu_1$, $\mathcal{H}_3$ admits of  a bound state  $| \omega \rangle$, given by
\begin{equation}
\omega_n= \frac{\cos [q_0(n-1)+ \sigma]}{\sqrt{\rho_n \rho_{n-1}}}.
\end{equation}

\begin{figure}[htbp]
  \includegraphics[width=84mm]{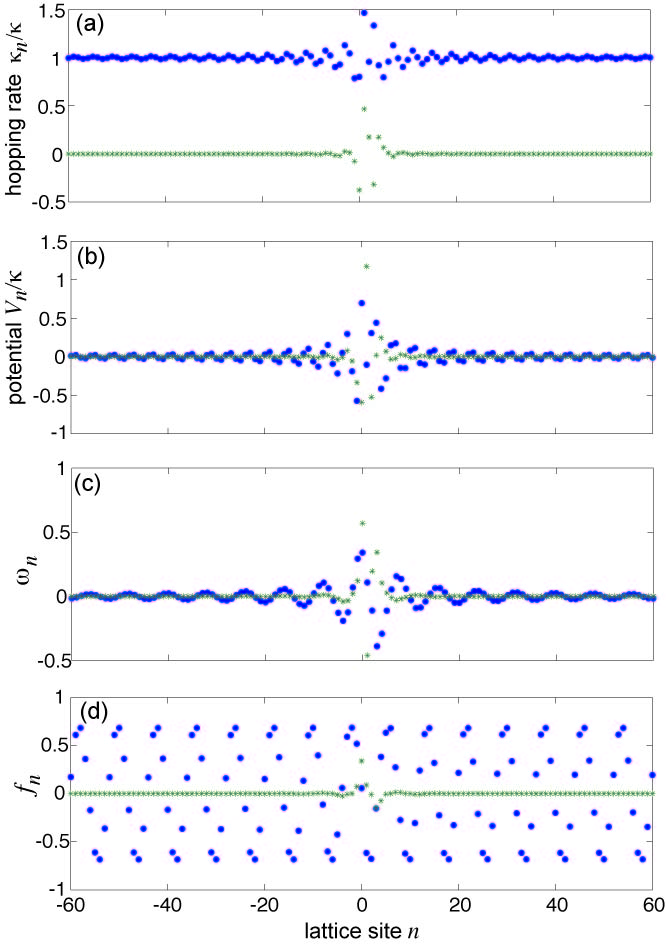}\\
   \caption{(color online) An example of a non-Hermitian lattice $\mathcal{H}_3$ synthesized by a double Darboux transformation for parameter values $\lambda=i$, $q_0=\pi/4$ and $\sigma=\pi/3$. (a) Behavior of the normalized hopping rates $\kappa_n/\kappa$. (b) Behavior of the normalized optical potential $V_n / \kappa$. (c) BIC mode amplitudes $\omega_n$ and (d) the corresponding associated function $f_n$. Bold dots and stars refer to the real and imaginary parts, respectively.}
\end{figure}

\subsection{Lattice properties}
The properties of the lattice described by the partner Hamiltonian $\mathcal{H}_3$ can be readily obtained from an analysis of Eqs.(3-6). In particular:\\
(i) Since $\rho_n \rightarrow \infty$ as $n \rightarrow \pm \infty$, one has $\kappa_n^{(3)} \rightarrow \kappa$ and $V_n^{(3)} \rightarrow 0$ as $n \rightarrow \pm \infty$, i.e. 
the lattice described by the Hamiltonian $\mathcal{H}_3$ is asymptotically an homogenous lattice. Physically, this means that we are dealing with an homogenous lattice with some {\it defects}. Moreover, since $\kappa_n^{(3)}$ and $V_n^{(3)}$ are generally complex-valued, the lattice is non-Hermitian (though rather generally $\mathcal{H}_3$ is not $\mathcal{PT}$ symmetric). An example of distributions of lattice hopping rates and site energy  potentials, synthesized by the double discrete Darboux transformation, is shown in Figs.1(a) and (b). \par
(ii) $\omega_n$ is vanishing like $ \sim 1 / n$ as $n \rightarrow \infty$, i.e. $\sum_n | \omega_n|^2 < \infty$ and thus the energy $E=\mu_1=2 \kappa \cos q_0$ belongs to the point spectrum of $\mathcal{H}_3$. Hence $| \omega \rangle$, defined by Eq.(6), is a BIC state for the lattice Hamiltonian $\mathcal{H}_3$. An example of the BIC mode distribution is shown in Fig.1(c).\par
(iii) The continuous spectrum of $\mathcal{H}_3$ is the energy interval $ - 2 \kappa \leq E \leq 2 \kappa$, with $E \neq \mu_1$.  For any energy $E= \mu$ in such an interval, with $\mu \neq \mu_1$, there are two non-normalizable (but limited) linearly-independent eigenfunctions $\xi_n^{\pm}$ of $\mathcal{H}_3$, which can be readily obtained from the plane-wave (Bloch) eigenstates $ \psi_n (q)\sim \exp( \pm iq n)$ of $\mathcal{H}_1$ via the linear transformation (A22) given in the Appendix A. They read explicitly
 \begin{equation}
 \xi_n^{\pm}(q)=\frac{ \kappa \exp(\pm i q n)}{\mu-\mu_1} \left[ A_n +B_n \exp( \mp i q)+C_n \exp(\mp 2  i q) \right]
 \end{equation} 
  where $\mu=2 \kappa \cos q$, 
  \begin{eqnarray}
  A_n & = & \sqrt{\frac{\rho_{n-1}}{\rho_n}} \\
  B_n & = &- \frac{\cos(q_0n+\sigma)}{\cos[q_0(n-1)+\sigma]} \sqrt{\frac{\rho_{n-1}}{\rho_n}} \nonumber \\
  & - & \frac{\cos[q_0(n-2)+\sigma]}{\cos[q_0(n-1)+\sigma]} \sqrt{\frac{\rho_{n}}{\rho_{n-1}}}   \\
  C_n & = &  \sqrt{\frac{\rho_{n}}{\rho_{n-1}}}
  \end{eqnarray}
and $\rho_n$ are given by Eq.(5). The asymptotic behavior of the solutions $\xi_{n}^{\pm }(q)$ as $n \rightarrow \pm \infty$ reads
\begin{equation}
\xi_n^{\pm} \simeq \exp[ \pm i q (n-1)]
\end{equation}   
from which it follows that the defect of the asymptotically-homogenous lattice defined by the Hamiltonian $\mathcal{H}_3$ is {\it invisible}, i.e. the transmission coefficient $t(q)$ for any incident Bloch wave, with wave number $q \neq \pm q_0$, is equal to one [$t(q)=1$ for $q \neq \pm q_0$].\par
(iv) The energy $E=\mu_1=2 \kappa \cos q_0$ is an EP in the continuum of $\mathcal{H}_3$ with $N=2$ algebraic multiplicity. This means  \cite{B2} that there exists an {\it associated function} $f_n$ to the BIC eigenstate $\omega_n$, which is a not-normalizable but limited function as $n \rightarrow \infty$ \cite{note}, such that 
\begin{eqnarray}
\left( \mathcal{H}_3 - \mu_1 \right) | \omega \rangle & = & 0 \\
\left( \mathcal{H}_3- \mu_1 \right)  | f \rangle & = & | \omega \rangle.
\end{eqnarray}   
The explicit expression of the associated function $|f \rangle$ is derived in the Appendix B and reads 
\begin{equation}
f_n=-\frac{i}{8 \kappa \sin^2 q_0} \lim_{q \rightarrow q_0} \frac{ \partial G_n} { \partial q}
\end{equation}
 where we have set
\begin{eqnarray}
G_n(q)  & = &  \exp(iqn+i \sigma) [A_n+B_n \exp(-iq)+C_n \exp(-2i q)] \;\;\;\;\; \nonumber \\
& - & \exp(-iqn-i \sigma)[A_n+B_n \exp(iq)+C_n \exp(2iq)].  \;\;\;\;\;\;\;\;
\end{eqnarray}
$f_n$ turns out to have the following asymptotic behavior as $n \rightarrow \pm \infty$ (see the Appendix B)
\begin{equation}
f_n \simeq -\frac{1}{2 \kappa \sin q_0} \sin [q_0(n-1)+\sigma].
\end{equation}
An example of the associate function $f_n$ to the BIC mode is shown in Fig.1(d).

\subsection{Exceptional Points and Bound States in the Continuum of von Neumann-Wigner type}
As shown in the previous subsection, the energy $E=\mu_1$ embedded into the continuous lattice spectrum belongs to the point spectrum of $\mathcal{H}_3$ and the normalizable state $ | \omega \rangle$ is effectively a BIC mode of the von Neumann-Wigner type \cite{Wigner}. However, there is a deep physical difference between the properties of an ordinary BIC mode of von Neumann-Wigner type in Hermitian systems and an EP in the continuum. To clarify this point, let us notice that an ordinary BIC state $| \omega \rangle$ in an Hermitian system is a {\it marginally stable state}, i.e. a perturbation added to $| \omega \rangle$ does not secularly grow. Conversely, the BIC mode $| \omega \rangle$ in a NHH which is an exceptional point in the continuum turns out to be an {\it unstable} state, even though the spectrum of the NHH is entirely real-valued. This means that a perturbation can induce a secular growth of the amplitude of the BIC mode $| \omega \rangle$, a feature which is a clear signature of the 'defective' nature of the EP. To show the unstable behavior of the BIC at an EP, let us note that the Schr\"{o}dinger equation
\begin{equation}
i \frac{\partial | \psi(t)  \rangle}{\partial t}= \mathcal{H}_3 | \psi (t) \rangle
\end{equation}
is satisfied by the function
\begin{equation}
| \psi(t) \rangle= \left[ (1-i \epsilon t)| \omega \rangle+ \epsilon |f \rangle \right] \exp(-i \mu_1 t)
\end{equation}
for any arbitrary value of the constant $\epsilon$, where $|f \rangle$ is the associated function to the BIC mode $| \omega \rangle$. This means that any arbitrarily small perturbation  shaped like the associated function $f_n$ will lead to a secular growth of the amplitude of the BIC mode. Correspondingly, the norm $\sqrt{|\langle \psi (t) | \psi(t) \rangle|}$ of the wave function will grow  linearly with time $t$, in spite the Hamiltonian has an entirely real energy spectrum. An example will be discussed in more details in the next section.\\ 
It should be finally noticed that, like for the continuous Schr\"{o}dinger equation \cite{B2}, the presence of an exceptional point in the continuum makes the mathematical problem of the resolution of the identity for the Hamiltonian $\mathcal{H}_3$ rather sophisticated. This nontrivial problem, however, will not be  considered in the present work.

\section{A simple $\mathcal{PT}$-symmetric optical lattice with an exceptional point in the continuum}
In this section we present a rather simple example of a NHH lattice that can describe light transport in arrays of evanescently-coupled optical waveguides, and discuss the physical properties of the EP in the continuum. The lattice is obtained using the double discrete Darboux transformation, outlined in the previous section, in the limiting case $\lambda=1$, $\sigma= \pi / 2$ and $q_0 \rightarrow \pi /2$. For such parameter values, from Eqs.(3-5) one obtains the isospectral lattice described by the Hamiltonian $\mathcal{H}_3$ with
\begin{equation}
\kappa_n^{(3)}=\kappa \left\{ 
\begin{array}{cc}
\sqrt{(n+1)/(n-1)} & n \; \; \rm {even} \\
\sqrt{(n-2)/n} & n \: \: \rm {odd}
\end{array}
\right.
\end{equation}
\begin{equation}
V_n^{(3)}=0.
\end{equation}
\begin{figure}[htbp]
  \includegraphics[width=85mm]{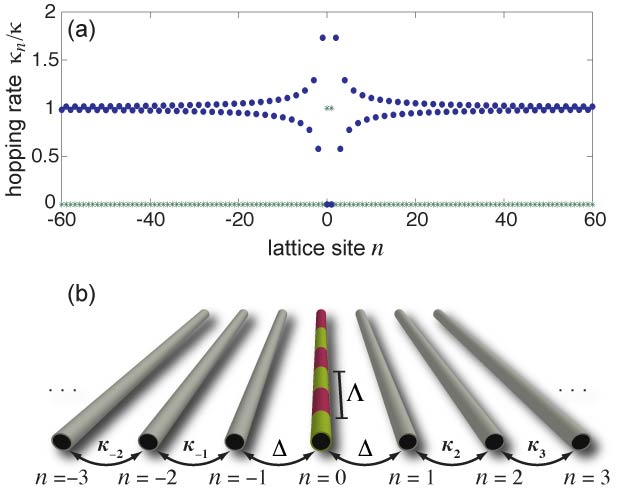}\\
   \caption{(color online)(a) Distribution of the normalized hopping rates $\kappa_n/ \kappa$ for the lattice defined by Eq.(19).  Bold dots and stars refer to the real and imaginary parts, respectively. 
(b) Schematic of an array of coupled optical waveguides that realizes the inhomogeneous hopping rates  in (a) with effective complex couplings at sites $n=-1,0$ and 1.}
\end{figure}
\begin{figure}[htbp]
  \includegraphics[width=85mm]{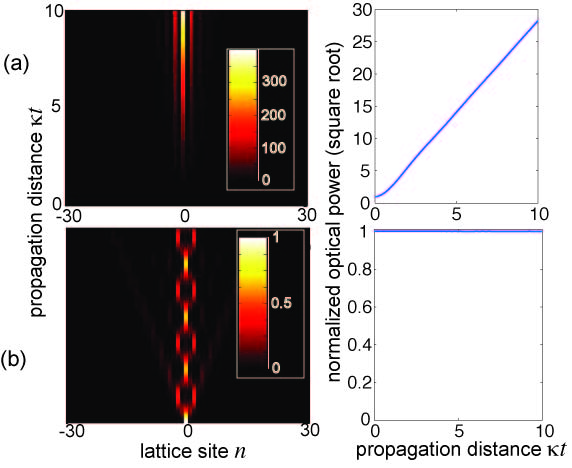}\\
   \caption{(color online) Beam propagation along (a) the $\mathcal{PT}$-symmetric lattice $\mathcal{H}_3$, and (b) the Hermitian lattice $\mathcal{H}_4$ for single waveguide excitation $c_n(0)=\delta_{n,0}$. The left panels show the evolution of the beam intensity $|c_n(t)|^2$ in a pseudocolor map, whereas the right panels show the evolution of the square root of normalized optical power, i.e. $\sqrt{P(t)/P(0)}$ with $P(t)=\sum_n |c_n(t)|^2$.}
\end{figure}
Note that the lattice has a inhomogeneous distribution of the hopping rates $\kappa_n^{(3)}$ in the neighborhood of the defective region (near $n=0$), whereas the site energies are homogeneus like for the initial lattice $\mathcal{H}_1$ (i.e. $V_n^{(3)}=V_n^{(1)}=0$). The distribution of the inhomogenous hopping rates is shown in Fig.2(a).  The non-Hemitian nature of the lattice comes from the fact that the hopping rates $\kappa_0$ and $\kappa_1$ are imaginary, namely $\kappa_0=\kappa_1=  \sqrt{-1} \kappa = \pm i \kappa$; for $n \neq 0,1$, the hopping rates $\kappa_n^{(3)}$ are instead real-valued. Note that there is some arbitrariness at this stage in the choice of the sign of $\kappa_0$ and $\kappa_1$, i.e. one can take $\kappa_0=-\kappa_1=i \kappa$ or $\kappa_0=\kappa_1=i \kappa$. The two choices, however, are essentially equivalent, since one can switch from one to the other by application of a $ \pi$ phase slip to the amplitudes $c_n$ above (or below) the $n=0$ site. We will consider here the  case $\kappa_0=-\kappa_1=i \kappa$, for which the lattice turns out to be $\mathcal{PT}$ symmetric. In coupled optical waveguides, an effective imaginary coupling constant can be realized by suitable longitudinal modulation of complex refractive index in the central waveguide $n=0$ of the lattice, as discussed in details at the end of this section (see also \cite{Darb2}). The inhomogeneous coupling constants $\kappa_n$  for $n \neq 0,1$ can be readily obtained by judicious waveguide spacing, as demonstrated for instance in the experiment of Ref.\cite{palle1bis}. A schematic of the optical waveguide array is shown in Fig.2(b).
The EP in the continuum occurs at the energy $\mu_1=2 \kappa \cos q_0=0$. The BIC mode $\omega_n$ and corresponding associated function $f_n$, as obtained from Eqs.(6) and (16), read explicitly
\begin{equation}
\omega_n = \left\{ 
\begin{array}{cc}
\frac{n}{|n|}\frac{i^n}{\sqrt{n^2-1}} & n \; \; \rm {even} \\
0 & n \: \: \rm {odd}
\end{array}
\right.
\end{equation}
\begin{equation}
f_n=-\frac{1}{2 \kappa} \sin ( \pi n /2).
\end{equation}
Note that the BIC mode $\omega_n$ has an algebraic decay and is similar to the von Neumann-Wigner BIC mode recently predicted and observed in Ref.\cite{palle1bis} for an Hermitian lattice with inhomogeneous hopping rates. However, as briefly mentioned in the previous section and discussed now in more details, the scattering and dynamical properties of a BIC at an EP deeply deviate from those of an ordinary BIC state in an Hermitian lattice. To clarify such a point, let us consider an associated {\it Hermitian} optical lattice defined by the Hamiltonian $\mathcal{H}_4$ with
\begin{equation}
\kappa_n^{(4)}=\kappa \left\{ 
\begin{array}{cc}
\sqrt{(n+1)/(n-1)} & n \; \; {\rm even}, \; n \neq 0 \\
\sqrt{(n-2)/n} & n \: \: {\rm odd}, \; n \neq 1 \\
1 & n=0 \\
-1 & n=1
\end{array}
\right.
\end{equation}
\begin{equation}
V_n^{(4)}=0.
\end{equation}
Basically, the Hermitian optical lattice $\mathcal{H}_4$ is obtained from the $\mathcal{PT}$-symmetric lattice  $\mathcal{H}_3$ by replacing the imaginary hopping rates $\kappa_0^{(3)}=-\kappa_1^{(3)}= i \kappa$ with the real ones $\kappa_0^{(4)}=-\kappa_1^{(4)}=\kappa$. It can be readily shown that the Hermitian lattice  $\mathcal{H}_4$ sustains a von Neumann-Wigner BIC mode at energy $E=0$, given by
\begin{equation}
\omega_n^{(H)} = \left\{ 
\begin{array}{cc}
\frac{n}{|n|}\frac{i^n}{\sqrt{n^2-1}} & n \; \; {\rm even}, \;\; n \neq 0 \\
1 & n=0 \\
0 & n \: \: \rm {odd}
\end{array}
\right.
\end{equation}
Note that such a BIC mode simply deviates from the BIC mode $\omega_n$ of $\mathcal{H}_3$ because of the different amplitude at lattice site $n=0$ [compare Eqs.(21) and (25)]. Obviously, $E=0$ is not an EP for $\mathcal{H}_4$, because  $\mathcal{H}_4$ is Hermitian. In addition to the BIC mode, the Hermtian lattice $\mathcal{H}_4$ sustans other two bound states with energy $E \simeq \pm 2.31 \kappa$ outside the lattice band, i.e. ordinary bound states in the gap (like those discussed in Ref.\cite{palle1bis}). To highlight the different physical properties between the BIC mode at the EP point for  the non-Hermitian lattice $\mathcal{H}_3$ and the ordinary BIC mode for the Hermitian lattice $\mathcal{H}_4$, we compare the propagation and scattering features of the two lattices.\\ 
{\it Propagation: single-site excitation}. In Fig.3 we show the numerically-computed evolution of the optical light intensity $|c_n(t)|^2$ for excitation of the $n=0$ lattice waveguide, i.e. for the initial condition $c_n(0)=\delta_{n,0}$, where in the optical context $t$ is an effective propagation distance. In the figure, the evolution of the square root of the normalized total optical power 
$\sqrt{P(t)/P(0)}$, with $P(t)=\sum_n |c_n(t)|^2$, is also shown. Note that, owing to the existence of the BIC mode, localization is observed in both the Hermitian and $\mathcal{PT}$-symmetric optical lattices. In the Hermitian case, a clear mode beating is observed, which arises from the excitation of the BIC mode and the other bound states in the gap; this scenario is similar to the one observed in the experiment of Ref. \cite{palle1bis}. Obviously the total optical power is conserved in this case. Conversely, for the $\mathcal{PT}$-symmetric lattice  [Fig.3(a)] the optical power shows a secular growth with time, in spite of the entire real-valued energy spectrum. As discussed in Sec.III.C, such an algebraic growth ($P(t) \sim t^2$)  is the clear signature of the instability of the BIC state because $E=0$ is an EP in the continuum.\\
{\it Scattering of plane waves}. To compare the scattering properties of the two optical lattices, we assume that the lattices are effectively homogeneous far from the defective region, i.e. we take $k_n=\kappa$ for $n \leq -N$ and $n \geq N+1$, where $N$ is a large enough integer \cite{notauffa}. In this case, the scattering states of the lattice in the homogenous regions $n \leq -N$ and $n \geq N$ are Bloch states with wave number $q$, corresponding to the energy $E=2 \kappa \cos q$. For a wave incident from the left side, the scattered state has the form
\begin{equation}
c_n(E)=\left\{ 
\begin{array}{cc}
\exp(-iqn)+r(q) \exp(iqn) &\\
&  n \leq -N \\
t(q) \exp(-iqn) & n \geq N
\end{array}
\right.
\end{equation}
 where $t(q)$ and $r(q)$ are the spectral transmission and reflection (for left-side incidence) coefficients. An accurate computation of $t(q)$ and $r(q)$ can be accomplished by a standard transfer matrix method, as discussed in the Appendix C. In Fig.4 we compare the spectral transmission and reflection coefficients for the two lattices $\mathcal{H}_3$ and $\mathcal{H}_4$, assuming $N=200$. Note that, according to the theoretical analysis of Sec.III.B,  the defect in the non-Hermitian lattice $\mathcal{H}_3$ is effectively invisible for $q \neq \pi/2$, i.e. $r(q)=0$ and $t(q)=1$ like in an effective homogeneous lattice. Near $q= \pi/2$, a narrow structured resonance deep (peak) in the spectral transmittance (reflectance) is observed, whose width shrinks as the number $N$ is increased. Conversely, the defect in the Hermitian lattice $\mathcal{H}_4$ is not manifestly an invisible defect. 
\begin{figure}[htbp]
  \includegraphics[width=85mm]{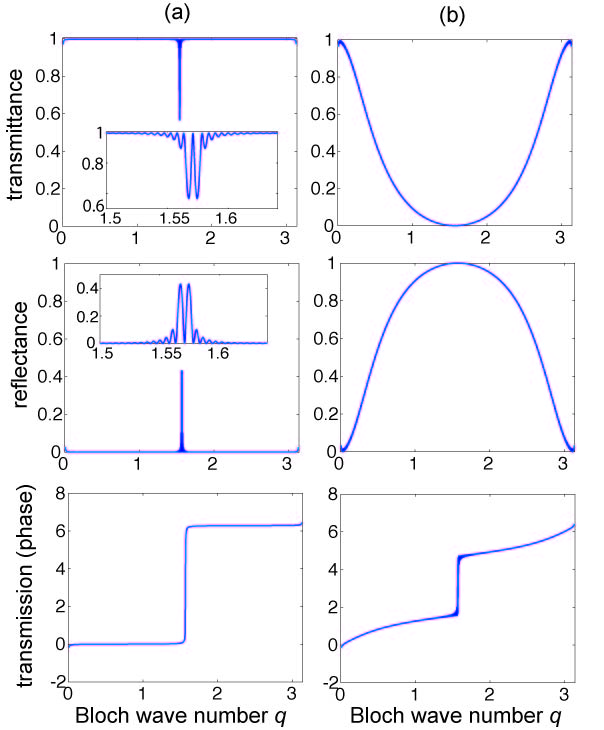}\\
   \caption{(color online) (a) Behavior of the numerically-computed spectral transmittance $|t|^2$ (upper panel), spectral reflectance $|r|^2$ (middle panel) and phase of the spectral transmission coefficient $t$ (lower panel) as a function of the Bloch wave number $q$ of incident wave for the $\mathcal{PT}$-symmetric optical lattice described by the Hamiltonian $\mathcal{H}_3$ [Eqs.(19,20)]. The insets in the upper and middle panels show an enlargement of the spectral behavior near the wave number $q= \pi/2$. (b) Same as (a), but for the Hermitian optical lattice defined by the Hamiltonian $\mathcal{H}_4$ [Eqs.(23,24)].}
\end{figure}

\par Finally, let us discuss a possible method to realize effective complex hopping rates
$\kappa_0^{(3)}=\kappa_1^{(3)}=i \kappa$ in the optical lattice. To this aim, let us consider an array of evanescently-coupled optical waveguides with engineered hopping rates $\kappa_n=\kappa_n^{(3)}$ as given by Eq.(19), except for $n \neq 0,1$ where we take $\kappa_0=\kappa_1=\Delta$,  $\Delta$ being a real-valued (Hermitian) coupling to be determined. In practice, inhomogeneous values of the couplings are realized by controlling the waveguide spacing \cite{palle1bis}. At the central waveguide $n=0$, we superimpose a longitudinal modulation of both effective propagation constant and optical gain/loss, described by a complex periodic function $\gamma(t)$ of the propagation distance $t$ with spatial period $\Lambda$; see Fig.2(b) for a schematic of the optical structure. Light propagation in such an array of waveguides is described by the set of coupled-mode equation 
\begin{equation}
i \frac{da_n}{dt}= \kappa_n a_{n-1}+\kappa_{n+1} a_{n+1}+ \gamma(t) \delta_{n,0} a_n
\end{equation}
for the modal amplitudes $a_n(t)$ of light trapped in the various guides. After setting $a_n(t)=c_n(t)$ for $n \neq 0$ and $a_0(t)=c_0(t)  \exp[-i \int_0^t d \xi \gamma(\xi)]$, 
for a rapidly oscillating function $\gamma (t)$, i.e. for $\Lambda$ smaller than $\sim 2 \pi / \kappa$, the rotating-wave approximation can be applied, leading to the following set of effective coupled-mode equations for the amplitudes $c_n(z)$ \cite{Darb2}
\begin{eqnarray}
i \frac{dc_n}{dt}=\kappa_n c_{n-1}+\kappa_{n+1}c_{n-1} \; \; n \neq 0, \pm1 \\
i \frac{dc_0}{dt}=\Delta R_+ ( c_{-1}+ c_{1})  \\
i \frac{dc_1}{dt}=\Delta R_- c_{0}+ \kappa_2 c_{2} \\
i \frac{dc_{-1}}{dt}=\Delta R_- c_{0}+ \kappa_{-1} c_{-2}
\end{eqnarray}
where we have set
\begin{equation}
R_{\pm} \equiv \frac{1}{\Lambda} \int_0^{\Lambda} dt \exp \left[ \pm i \int_0^t d \xi \gamma(\xi) \right]
\end{equation}
The lattice with effective hopping rates (19) is thus obtained provided that the longitudinal modulation $\gamma(t)$ of the complex refractive index in waveguide $n=0$ is chosen to satisfy the constraint
\begin{equation}
R_+=R_-=i \kappa / \Delta.
\end{equation}
This condition can be realized for a wide range of modulation profiles. Let us discuss two possible cases.\\
(i) {\it Sinusoidal modulation.} Let us assume a modulation of the effective refractive index of the form
\begin{equation}
\gamma(t)=(\alpha+i \beta) \cos (2 \pi t / \Lambda)
\end{equation}
where $\alpha$ and $\beta$ are the modulation depths of the real (propagation constant detuning) and imaginary (gain/loss term) parts of the modulation. In this case one has $R_{\pm}=J_0(\Gamma)$, where $\Gamma=\Lambda (\alpha+i \beta)/ (2 \pi)$ and $J_0$ is the Bessel function of first kind and zero order. If, for instance, we assume $\alpha=2 \times 2 \pi / \Lambda$ and $\beta=-2.096 \times 2 \pi / \Lambda$, one has $R_{\pm}=J_0(\Gamma) \simeq 1.9414 i$ and the condition (33) is thus satisfied for $\Delta \simeq \kappa/1.9414 $.\\
(i) {\it Square-wave modulation.} Let us assume a square-wave modulation of the effective refractive index of the form
\begin{equation}
\gamma(t)=(\alpha+i \beta) H (2 \pi t / \Lambda)
\end{equation}
where $H(x)$ is a $2 \pi$-periodic square wave with zero mean, i.e. $H(x+2 \pi)=H(x)$, $H(x)=1$ for $0<x< \pi/2$ and $3 \pi/2< x<2 \pi$, and $H(x)=-1$ for $\pi/2<x< 3 \pi/2$. This means that the central waveguide $n=0$ is segmented,  with segments of lengths $\Lambda /2$ that alternate optical amplification/loss  and high/low effective index.
In this case one has $R_{\pm}=\sin( \Gamma) / \Gamma$, where now $\Gamma= \Lambda (\alpha+i\beta)/4$. If, for instance, we assume $\alpha=2.7255 \times 4 / \Lambda$ and $\beta=-1.3707 \times 4 / \Lambda$, one has $R_{\pm}=\sin(\Gamma)/ \Gamma \simeq 0.6182 i$ and the condition (33) is thus satisfied for $\Delta \simeq \kappa/0.6182$.\\

\section{Conclusions}
The spectral and dynamical properties of non-Hermitian (including $\mathcal{PT}$-symmetric) systems are strongly influenced by the appearance of exceptional points and spectral singularities. In Hamiltonians with an entire real energy spectrum, such singular energies usually appear at the onset of $\mathcal{PT}$ symmetry breaking and are responsible for a secular (unstable) growth of the wave function in spite of the reality of the energy spectrum. Exceptional points are generally found in finite dimensional Hamiltonians with a discrete energy spectrum, whereas spectral singularities are defective states belonging to the continuous energy spectrum and thus can not be found in finite-dimensional systems. 
In this work we have shown that a novel class of exceptional points, namely exceptional points in the continuum, can arise in non-Hermitian optical lattices with engineered defects. At an exceptional point, the lattice sustains a bound state with an energy embedded in the spectrum of scattered states, similar to the von Neumann-Wigner bound states in the continuum of Hermitian lattices. Such states can be sustained in defective lattices synthesized by application of a double discrete Darboux (supersymmetric) transformation to the homogeneous Hermitian lattice. The dynamical and scattering properties of bound states in the continuum  at an exceptional point are deeply modified by the defective nature of an exceptional point. In particular, contrary to the usual von Neumann-Wigner bound states of Hermitian systems, the bound states in the continuum at an exceptional point  are unstable states that can secularly grow by an infinitesimal perturbation. Such properties have been discussed in details for transport of discretized light in a $\mathcal{PT}$-symmetric array of coupled optical waveguides, which could provide an experimentally accessible system to observe exceptional points in the continuum.
\acknowledgments
This work was supported by the
Fondazione Cariplo (Project ÒNew Frontiers in Plasmonic
NanosensingÓ, Rif. 2011-0338). 

\appendix

\section{Double Draboux transformation for the discrete Schr\"{o}dinger equation}
In this Appendix we provide, for the sake of completeness and clearness of the analysis, a brief review of the single and double Darboux transformation techniques for the discrete 
Schr\"{o}dinger equation. For a more comprehensive and extended study of the discrete Darboux technique we refer the reader to previous references \cite{Darb1,Darb2}.

\subsection{Simple Darboux transformation}
Let us consider a one-dimensional tight-binding lattice described by
the Hamiltonian
\begin{equation}
\mathcal{H}= \sum_{n}  \kappa_{n} \left( |n-1\rangle \langle n|+|n
\rangle \langle n-1|\right)  + \sum_n V_n |n \rangle \langle n|
\end{equation}
where $|n\rangle$ is a Wannier state localized at site $n$ of the
lattice, $\kappa_n$ is the hopping rate between sites $|n-1 \rangle$
and $|n \rangle$, and $V_n$ is the energy of Wannier state $|n
\rangle$. The energy spectrum $E$ of $\mathcal{H}$ is obtained from the eigenvalue problem of the discrete
Schr\"{o}dinger equation
\begin{equation}
\kappa_{n} \psi_{n-1}+\kappa_{n+1} \psi_{n+1}+V_n \psi_{n}=E \psi_n
\end{equation}
with eigenstate $|\psi \rangle= \sum_n \psi_n |n \rangle$. The point spectrum of $\mathcal{H}$ corresponds to energies $E$ with normalizable eigenstates ($\sum_n |\psi_n|^2 < \infty$), whereas the
continuous spectrum of $\mathcal{H}$ corresponds to to energies $E$ at which the eigenstate is not normalizable but $|\psi_n|$ is limited as $n \rightarrow \pm \infty$ (improper eigenfunctions). 
For instance, for the homogeneous Hermitian lattice, corresponding to $\kappa_n=\kappa$ and $V_n=0$, the energy spectrum is purely continuous and given by the usual tight-binding band $- 2 \kappa \leq E \leq 2 \kappa$ with (improper) eigenstates $\psi_n=\exp(iqn)$, where $-\pi \leq q < \pi$ is the Bloch wave number and $E=2 \kappa \cos q$. Note that $\mathcal{H}$ turns out to be Hermitian provided
that the hopping amplitudes $\kappa_n$ and site energies $V_n$ are
 real-valued parameters.\\ Let us indicate by $\mathcal{H}_1$ the tight-binding Hamiltonian
defined by Eq.(A1) with hopping amplitudes and site energies given by
$\kappa^{(1)}_n$ and $V^{(1)}_n $, respectively, and let us assume
that $\kappa_n^{(1)} \rightarrow \kappa>0$ and $V_n^{(1)}
\rightarrow 0$ as $n \rightarrow \pm \infty$, i.e. that the lattice
is asymptotically homogeneous. Let  us then 
indicate by $| \phi^{(1)} \rangle= \sum_n \phi_n^{(1)} | n \rangle$
one of the two linearly-independent solutions to the discrete Schr\"{o}dinger equation
$\mathcal{H}_1 | \phi^{(1)} \rangle=\mu_1  | \phi^{(1)} \rangle$, i.e.
\begin{equation}
\kappa_n^{(1)} \phi_{n-1}^{(1)}+\kappa_{n+1}^{(1)}
\phi_{n+1}^{(1)}+V^{(1)}_n \phi_{n}^{(1)}=\mu_1 \phi_{n}^{(1)}
\end{equation}
where $\mu_1$ can or can not belong to the spectrum of $\mathcal{H}_1$. 
It can be then shown that the following factorization for $\mathcal{H}_1$
holds \cite{Darb2}
\begin{equation}
\mathcal{H}_1=\mathcal{Q}_1 \mathcal{R}_1+\mu_1
\end{equation}
where
\begin{eqnarray}
\mathcal{Q}_1 & = & \sum_n \left( q_{n}^{(1)} |n \rangle \langle
n|+\bar{q}_{n-1}^{(1)} |n-1 \rangle \langle n| \right) \\
\mathcal{R}_1 & = & \sum_n \left( r_{n}^{(1)} |n \rangle \langle
n|+\bar{r}_{n+1}^{(1)} |n+1 \rangle \langle n| \right)
\end{eqnarray}
and
\begin{eqnarray}
r_n^{(1)} & = & - \sqrt{\frac{\kappa_n^{(1)}
\phi_{n-1}^{(1)}}{\phi_n^{(1)}}} \\
\bar{r}_n^{(1)} & = & -\frac{\kappa_{n}^{(1)}}{r_{n}^{(1)}} \\
q_n^{(1)} & = & -r_n^{(1)} \\
\bar{q}_n^{(1)} & = & - \bar{r}_{n+1}^{(1)}.
\end{eqnarray}
Let us then introduce the new Hamiltonian $\mathcal{H}_2$ obtained
from $\mathcal{H}_1$ by interchanging the operators $\mathcal{R}_1$
and $\mathcal{Q}_1$, i.e. let us set
\begin{equation}
\mathcal{H}_2=\mathcal{R}_1 \mathcal{Q}_1+\mu_1.
\end{equation}
$\mathcal{H}_2$ will be referred to as the partner Hamiltonian of
$\mathcal{H}_1$. $\mathcal{H}_2$ describes the Hamiltonian of a
tight-binding lattice [i.e., it is of the form (A1)] with hopping
amplitudes and site energies $\{ \kappa^{(2)}_{n},V^{(2)}_n\}$ given
by 
\begin{eqnarray}
\kappa^{(2)}_n & = & \kappa_n^{(1)} \frac{r^{(1)}_{n-1}}{r^{(1)}_n} \\
V_n^{(2)} & = & V_n^{(1)}+\kappa_{n+1}^{(1)}
\frac{\phi_{n+1}^{(1)}}{\phi_n^{(1)}}-\kappa_n^{(1)}
\frac{\phi^{(1)}_n}{\phi^{(1)}_{n-1}}.
\end{eqnarray}
Note that, to avoid the occurrence of divergences in $\kappa_{n}^{(2)}$ and $V_{n}^{(2)}$, the sequence $\phi_{n}^{(1)}$ is generally requested  not to vanish at any $n$. The following properties then hold:\\
(i) If $| \psi \rangle$ satisfies the discrete Schr\"{o}dinger equation $\mathcal{H}_1 | \psi \rangle= \mu | \psi \rangle$ with $ \mu \neq \mu_1$, then the state 
$|\xi \rangle = \mathcal{R}_1 | \psi \rangle$, i.e.
 \begin{equation}
\xi_n =r_n^{(1)} \psi_n+\bar{r}_n^{(1)} \psi_{n-1}
\end{equation}
satisfies the equation $\mathcal{H}_2 | \xi \rangle= \mu | \xi \rangle$.\\
(ii) The equation $\mathcal{H}_2 | \psi \rangle= \mu_1 | \psi \rangle$ is satisfied for $| \psi \rangle =| \theta \rangle$ with Wannier amplitudes
\begin{equation}
\theta_n=\frac{1}{\sqrt{\kappa_n^{(1)} \phi^{(1)}_n \phi^{(1)}_{n-1}
}}.
\end{equation}
 We remark that in the previous relations $\mu$ and $\mu_1$ can or cannot belong to the spectrum of $\mathcal{H}_1$. An important consequence of the two above properties is  that the two Hamiltonians $\mathcal{H}_1$ and $\mathcal{H}_2$ are isospectral, i.e. they have the same energy spectrum, apart form $E=\mu_1$, which might or might not belong  to the spectrum of either one of $\mathcal{H}_1$ or $\mathcal{H}_2$.

\subsection{Double Darboux transformation}
According to the analysis of the previous subsection, the function $| \theta \rangle = \sum_n \theta_n | n \rangle$ defined by Eq.(A15) is a solution to the discrete Schr\"{o}dinger equation $\mathcal{H}_2 | \theta \rangle= \mu_1 | \theta \rangle$. The most general solution $| \phi^{(2)} \rangle$ to the same equation can be readily calculated and, apart from an unessential multiplication constant, reads
\begin{equation}
\phi^{(2)}_n= \theta_n \left( \lambda+\sum_{k=0}^{n-1} \frac{1}{ \kappa^{(2)}_k \theta_k \theta_{k+1} } \right)
\end{equation}
where $\lambda$ is an arbitrary complex-valued number.  Besides the decomposition (A11), we can also formally write
\begin{equation}
\mathcal{H}_2={\mathcal Q}_2 {\mathcal R}_2+ \mu_1
\end{equation}
where we have set
\begin{eqnarray}
\mathcal{Q}_2 & = & \sum_n \left( q_{n}^{(2)} |n \rangle \langle
n|+\bar{q}_{n-1}^{(2)} |n-1 \rangle \langle n| \right) \\
\mathcal{R}_2 & = & \sum_n \left( r_{n}^{(2)} |n \rangle \langle
n|+\bar{r}_{n+1}^{(2)} |n+1 \rangle \langle n| \right)
\end{eqnarray}
The expressions of $r_n^{(2)}$, $\bar{r}_n^{(2)}$, $q_n^{(2)}$, $\bar{q}_n^{(2)}$ entering in Eqs.(A18) and (A19) have the same form as Eqs.(A7-A10), with $^{(1)}$ replaced by $^{(2)}$. We then introduce the new Hamiltonian $\mathcal{H}_3$ obtained
from $\mathcal{H}_2$ by interchanging the operators $\mathcal{R}_2$
and $\mathcal{Q}_2$, i.e. 
\begin{equation}
\mathcal{H}_3=\mathcal{R}_2 \mathcal{Q}_2+\mu_1.
\end{equation}
The Hamiltonian $\mathcal{H}_3$ describes  a
tight-binding lattice with hopping
amplitudes and site energies $\{ \kappa^{(3)}_{n},V^{(3)}_n\}$, which have
the same expressions as Eqs.(A12) and (A13) provided that 
the replacement $^{(1) \rightarrow (2)}$ is made on the right hand sides of the equations.
The Hamiltonian $\mathcal{H}_3$ is isospectral to $\mathcal{H}_1$, apart from the energy value $E=\mu_1$ which needs to be separately investigated. 
The state $| \omega \rangle$ defined by 
\begin{equation}
\omega_n=\frac{1}{\sqrt{\kappa_n^{(2)} \phi_n^{(2)} \phi_{n-1}^{(2)} }}
\end{equation}
satisfies the equation $\mathcal{H}_3 | \omega \rangle= \mu_1 | \omega \rangle$. For $\mu \neq \mu_1$, the solution $|\xi \rangle$ to the equation $\mathcal{H}_3 | \xi \rangle= \mu | \xi \rangle$  can be readily obtained from the solution $| \psi \rangle$ of the discrete Schr\"{o}dinger equation $\mathcal{H}_1 | \psi \rangle= \mu | \psi \rangle$ via the linear transformation $|\xi \rangle = \mathcal{R}_2 \mathcal{R}_1 | \psi \rangle$, i.e.
\begin{equation}
\xi_n= r_n^{(2)}r_n^{(1)} \psi_n+ [r_n^{(2)} \bar{r}_n^{(1)}+r_{n-1}^{(1)} \bar{r}_n^{(2)} ] \psi_{n-1}+\bar{r}_{n-1}^{(1)} \bar{r}_{n}^{(2)} \psi_{n-2}.
\end{equation}

\section{Associated function to the BIC mode}
In this Appendix we show that the energy $E=\mu_1$ is an exceptional point of the Hamiltonian $\mathcal{H}_3$ with algebraic multiplicity $N=2$, i.e. that 
there exists a non-normalizable but limited associated function $|f \rangle$ satisfying Eq.(13) given in the text. To this aim, let us consider the following linear combination of 
improper eigenfuctions of $\mathcal{H}_3$
\begin{equation}
F_n(q)=i \frac{\mu-\mu_1}{4 \kappa \sin q_0} \left[ \xi_n^+ (q) \exp(i \sigma)-\xi_n^- \exp(-i \sigma) \right]
\end{equation}
where $\mu=\mu(q)=2 \kappa \cos q$, $\mu_1=\mu(q_0)$, and $\xi_n^{\pm }(q)$ are defined by Eqs.(7-10) given in the text.  Contrary to $\xi_n^{\pm}(q)$ which are singular at $q=q_0$, $F_n(q)$ is a limited and non-singular function for any value of $q$, including $q=q_0$. In fact, after some tedious but straightforward algebra one can show that the BIC mode $\omega_n$ is obtained from $F_n(q)$ in the limit $q \rightarrow q_0$, i.e. 
\begin{equation}
 \omega_n=\lim_{q \rightarrow q_0}F_n(q).
 \end{equation} 
Let us introduce the shifted Hamiltonian $\mathcal{H}^{'}=\mathcal{H}_3-\mu_1$ and the function $\psi_n(q)=F_n(q)/(q-q_0)$. Hence $\mathcal{H}^{'}\psi_n(q)=(\mu-\mu_1) \psi_n(q)$ for any $q$ in the neighborhood of $q_1$, with $q \neq q_1$, and $\mu=\mu(q)$. Let us then apply the operator $\mathcal{H}^{'}$ to the  function $(q-q_0) (\partial \psi_n/ \partial q)$. One has
\begin{eqnarray}
{\mathcal H}^{'} \left( (q-q_0) \frac{\partial \psi_n}{\partial q} \right) & = & (q-q_0) \frac{\partial}{\partial q} \mathcal{H}^{'} \psi_n = \nonumber \\
= \frac{\partial \mu}{\partial q}F_n+ (\mu-\mu_1)\frac{\partial F_n}{\partial q}  & - & (\mu-\mu_1) \psi_n  
\end{eqnarray}
i.e.
\begin{eqnarray}
{\mathcal H}^{'} \left( \psi_n+ (q-q_0) \frac{\partial \psi_n}{\partial q}  \right)  = \nonumber \\
= \frac{\partial \mu}{\partial q}F_n+ (\mu-\mu_1)\frac{\partial F_n}{\partial q} . 
\end{eqnarray}
Since $\psi_n+(q-q_0) (\partial \psi_n/ \partial q)=(\partial F_n/ \partial q)$, from Eq.(B4) one obtains
\begin{equation}
\mathcal{H}^{'} \frac{\partial F_n}{\partial q}=\frac{\partial \mu}{\partial q}F_n+ (\mu-\mu_1)\frac{\partial F_n}{\partial q}.
\end{equation}
If we take the limit of both sides in Eq.(B5) for $q \rightarrow q_0$, since $\mu-\mu_1 \rightarrow 0$, $F_n(q) \rightarrow \omega_n$ and $\partial F_n/ \partial q$ is a non-singular function, one has
\begin{equation}
\mathcal{H}^{'} \left( \lim_{q \rightarrow q_0} \frac{\partial F_n}{\partial q} \right)=\left( \frac{\partial \mu}{\partial q} 	\right)_{q_0} \omega_n
\end{equation}
Taking into account that $(\partial \mu / \partial q)_{q_0}=-2 \kappa \sin q_0$, $\mathcal{H}^{'}= \mathcal{H}_3-\mu_1$ and using Eqs.(7) and (B1), one finally obtains
\begin{equation}
\left( \mathcal{H}_3-\mu_1 \right) f_n= \omega_n
\end{equation}
where we have set
\begin{equation}
f_n= -\frac{i}{8 \kappa \sin^2 q_0 } \lim_{q \rightarrow q_0} \frac{ \partial G_n(q)}{\partial q}
\end{equation}
and $G_n(q)$ is defined by Eq.(15) given in the text. It should be noted that $(\partial G_n / \partial q)$ contains secularly growing terms with the power law $ \sim n$ as $n \rightarrow \infty$, however it can be readily shown from Eqs.(8,9,10) and (15) that such terms vanish when taking the limit  $q \rightarrow q_0$, i.e. $f_n$ is a limited function with respect to index $n$. More precisely, the following asymptotic behavior for the associated function $f_n$ as $ n \rightarrow \pm \infty$ can be readily obtained after taking the limit $ q \rightarrow q_0$ in Eq.(B8)
\begin{equation}
f_n \simeq -\frac{1}{2 \kappa \sin q_0} \sin [q_0(n-1)+\sigma]. 
\end{equation}

\section{Transfer matrix method for the computation of the spectral transmission and reflection coefficients of a defective lattice}
In this Appendix we briefly discuss an accurate transfer matrix method to compute the spectral transmission and reflection coefficients in a defective lattice, described by a rather generic Hamiltonian $\mathcal{H}$ given by Eq.(1) in the text, with  $\kappa_n \rightarrow \kappa$ and $V_n \rightarrow 0$ as $n \rightarrow \pm \infty$. To this aim, let us consider a sufficiently large integer $N$, and let us effectively assume $\kappa_n=\kappa$ for $n \leq -N$, $n \geq N+1$ and $V_n=0$ for $|n| \geq N$. This is a physically reasonable assumption, since in practice the defective region  can be considered compact. Under such an assumption, there are generally two linearly-independent scattered states of $\mathcal{H}$ with energy $E=2 \kappa \cos q$, corresponding to a Bloch wave incident from either the left or right side of the defect. For left-side incidence, in the homogenous lattice regions the scattered state has the form 
\begin{equation}
c_n(q)=\left\{ 
\begin{array}{cc}
\exp(-iqn)+r(q) \exp(iqn) &\\
&  n \leq -N \\
t(q) \exp(-iqn) & n \geq N
\end{array}
\right.
\end{equation}
 where $t(q)$ and $r(q)$ are the spectral transmission and reflection (for left-side incidence) coefficients, and $q$ is the Bloch wave number ($ 0 \leq q \leq \pi$). To determine the expressions of $t(q)$ and $r(q)$, let us note that from Eq.(A2) the following relation holds
 \begin{equation}
 \left( 
 \begin{array}{c}
 c_{n+1} \\
 c_{n}
\end{array} 
 \right)= \mathcal{M}_n
 \left( 
 \begin{array}{c}
 c_{n} \\
 c_{n-1}
\end{array}
\right)
\end{equation} 
 where we have set
 \begin{equation}
\mathcal{M}_n=\left( 
 \begin{array}{cc}
 (E-V_n)/ \kappa_{n+1} & -\kappa_n / \kappa_{n+1} \\
 1 & 0
\end{array}
\right)  
 \end{equation}
 and $E=2 \kappa \cos q$.
Hence one has
\begin{equation}
 \left( 
 \begin{array}{c}
 c_{N+1} \\
 c_{N}
\end{array} 
 \right)= \mathcal{Q}
 \left( 
 \begin{array}{c}
 c_{-N} \\
 c_{-N-1}
\end{array}
\right)
\end{equation} 
 where 
 \begin{equation}
  \mathcal{Q}=\mathcal{M}_N \times \mathcal{M}_{N-1} \times .... \times \mathcal{M}_{-N+1} \times \mathcal{M}_{-N}.
  \end{equation}
   From Eqs.(C1) and (C4) it then follows
\begin{eqnarray}
\left(
\begin{array}{c}
t(q) \exp[-iq(N+1)] \\
t(q) \exp(-iqN) 
\end{array}
\right)=
\left(
\begin{array}{cc}
\mathcal{Q}_{11} & \mathcal{Q}_{12} \\
\mathcal{Q}_{21}  & \mathcal{Q}_{22}
\end{array}
\right) \times \nonumber \\
\left( 
\begin{array}{c}
\exp(iqN)+r(q) \exp(-iqN) \\
\exp[iq(N+1)] +r(q) \exp[-iq(N+1)]
\end{array}
\right)
\end{eqnarray}   
where $\mathcal{Q}_{ik}$ are the elements of the $2 \times 2$ matrix $\mathcal{Q}$. Equation (C6)  can be solved, yielding the following expressions for the reflection and transmission coefficients
\begin{widetext}
 \begin{eqnarray}
 r(q)= \exp(2iqN) \frac{ \mathcal{Q}_{21} \exp(-iq)-\mathcal{Q}_{12} \exp(iq)+\mathcal{Q}_{22}-\mathcal{Q}_{11}}{\mathcal{Q}_{11}-\mathcal{Q}_{22} \exp(-2iq)+(\mathcal{Q}_{12}-\mathcal{Q}_{21}) \exp(-iq)} \;\;\;\;\;\;\; \\
 t(q)=\mathcal{Q}_{11} \exp(iq)[\exp(2iqN)+r(q)]+\mathcal{Q}_{12} [\exp(2iqN+2iq)+r(q)] \;\;\;\;\;\;\;\;\;\;
 \end{eqnarray}
 \end{widetext}
Hence, to compute the spectral transmission and reflection coefficients, for a fixed value of the Bloch wave number $q$ one first calculate the transfer matrix $\mathcal{Q}$ using Eqs.(C3) and (C5), and then one calculate $r(q)$ and $t(q)$ using Eqs.(C7) and (C8).


\begin{thebibliography}{31}
\bibitem{note}
This means that the associated function $f_n$ does not belong to the Hilbert space, however it has the same properties of the improper eigenfunctions of $\mathcal{H}_3$.
\end{thebibliography}

\end{document}